\documentclass[aps,twocolumn,showpacs,superscriptaddress,groupedaddress, epsf, nofootinbib]{revtex4}
\usepackage{graphicx, float}  
\usepackage{dcolumn}   
\usepackage{bm, amssymb, latexsym, amsfonts, amsmath}        
\usepackage{mathrsfs,array,multirow}
\usepackage[usenames,dvipsnames]{color}
\usepackage{xcolor}
\usepackage[colorlinks=true,linkcolor=blue]{hyperref}
\usepackage{epsfig, subfigure}
\usepackage{multirow}
\usepackage{rotating}
\usepackage{dcolumn}
\usepackage{soul}

\newcommand{\be}[1]{\begin{equation} \label{#1}}
\newcommand{\ee}{\end{equation}}
\newcommand{\bea}{\begin{eqnarray}}
\newcommand{\bean}{\begin{eqnarray*}}
\newcommand{\eea}{\end{eqnarray}}
\newcommand{\eean}{\end{eqnarray*}}
\newcommand{\ba}{\begin{array}}
\newcommand{\ea}{\end{array}}
\newcommand{\bel}{\begin{align}}
\newcommand{\eel}{\end{align}}

\begin{document}

\title{Charged wormholes in (anti-)de Sitter spacetime}

\author{Hyeong-Chan Kim}
\email{hckim@ut.ac.kr}
\affiliation{School of Liberal Arts and Sciences, Korea National University of Transportation, Chungju 27469, Korea}

\author{Wonwoo Lee}
\email{warrior@sogang.ac.kr}
\affiliation{Center for Quantum Spacetime, Sogang University, Seoul 04107, Korea}

\date{\today}

\begin{abstract}
We present a family of charged, traversable wormhole solutions in the presence of a cosmological constant.
In de Sitter spacetime, two types of wormhole throats can exist--referred to as typical and cosmological throat--located at small and large radial values, respectively.
In anti-de Sitter spacetime, the throat geometry allows for positive, zero, or negative curvature, enabling the possibility of an infinite throat area.
We analyze the flare-out condition, a key requirement for the existence of traversable wormholes, which imposes constraints on the equation of state parameters governing the supporting matter.
These solutions are shown to be of Petrov type $D$.
Furthermore, we examine radial geodesics of null and timelike particles.
In the de Sitter case, particles traverse the wormhole, passing from one throat to the other.
In contrast, in the anti-de Sitter case, particles exhibit recurrent oscillatory motion between two asymptotic regions, cyclically disappearing and reappearing across the throats.
\end{abstract}

\maketitle


\textit{Introduction ---} Wormholes are one of the most intriguing solutions allowed by general relativity.
Since the work of Einstein and Rosen, who first examined two asymptotically flat universes connected by a throat~\cite{Einstein:1935tc},
numerous studies have investigated solutions for traversable wormholes within the Einstein theory of gravitation,
as well as their characteristics~\cite{Misner:1957mt, Ellis:1973yv, Morris:1988cz, Morris:1988tu}.
These investigations have revealed that the matter necessary to sustain a wormhole must
violate the null energy condition near the throat and is often dubbed as
as exotic matter or a phantom field~\cite{Visser:1995cc, Lobo:2005us}.
Although such matter has yet to be observationally confirmed,
the vastness of the Universe suggests that its existence in some regions
could render the presence of wormholes plausible.
Driven by this consideration, some have also sought to understand how wormholes
if they exist, might exhibit observable effects distinct from those of black holes~\cite{Cramer:1994qj, Tsukamoto:2012xs, Bambi:2013nla, Nedkova:2013msa, DeFalco:2020afv, Godani:2021aub}.

Theoretically, wormhole solutions have been developed not only within various theories of gravitation~\cite{Kim:1997jf, Lobo:2007qi, Ovgun:2018xys, Maldacena:2018gjk, Halder:2019urh, Kim:2019ojs, Godani:2020gbr, Godani:2021egp, Godani:2022wkh, Godani:2023jyx, Battista:2024gud, Jang:2024nhm}
but also in gravitational theories that incorporate higher curvature terms~\cite{Lobo:2009ip, Kanti:2011jz}.
Studies have suggested that thanks to these modified gravity effects,
it is possible to construct wormholes without the direct necessity of a phantom field as a supportive matter.
Additionally, wormhole solutions have been examined in the presence of supplementary matter fields,
such as the Maxwell field~\cite{Kim:2001ri, Huang:2019arj, Kim:2024mam}.
Furthermore, research on wormholes that involve the Dirac field has been actively pursued,
yielding intriguing results~\cite{Blazquez-Salcedo:2020czn, Konoplya:2021hsm, Blazquez-Salcedo:2021udn}.
In a simplistic model, our Universe undergoes expansion driven by a cosmological constant, which also has room to evolve.
It would be fascinating to construct wormhole solutions that incorporate such a cosmological constant~\cite{Kim:2003zb, Kim:2016pky, Blazquez-Salcedo:2020nsa}.
In line with this motivation, wormhole solutions have been derived in the context of cosmological metrics~\cite{Roman:1992xj, Kim:1995xf}.
A thin-shell wormhole solution of the Schwarzschild-de Sitter type has been analyzed~\cite{Kim:1992sh, Lemos:2003jb, Lemos:2004vs}.
In this paper, we aim to obtain static traversable wormhole solutions in the presence of a cosmological constant by solving
the Einstein and Maxwell equations, naturally extending the concept of charge without charge described in Refs.~~\cite{Misner:1957mt, Kim:2024mam}.
\begin{figure}[h]
\begin{center}
\subfigure[wormhole in de Sitter]
{\includegraphics[width=1.08 in]{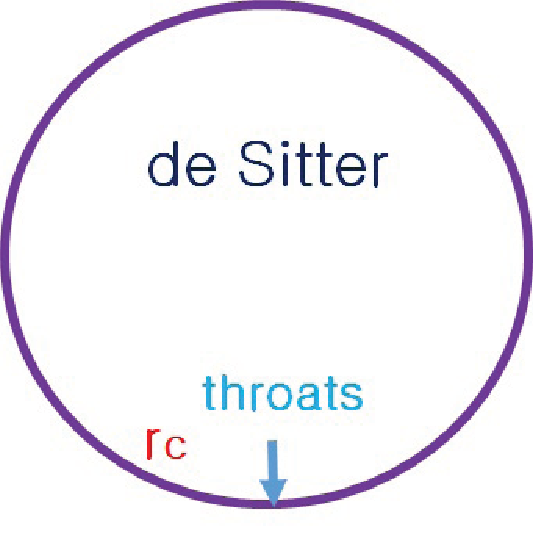}}
\subfigure[wormhole in de Sitter]
{\includegraphics[width=1.4 in]{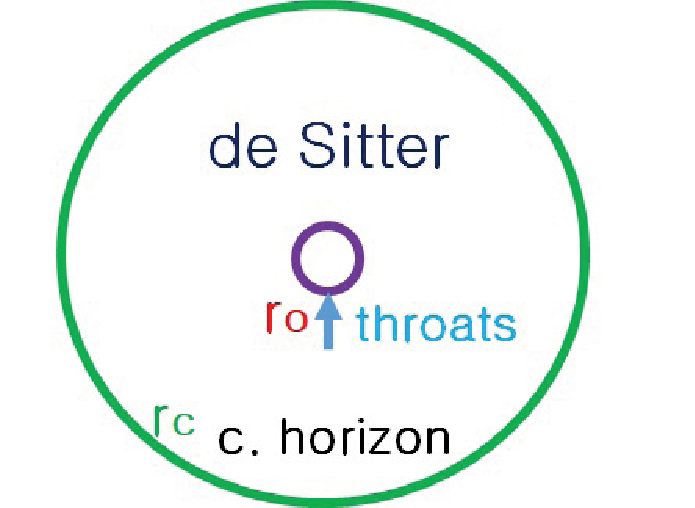}}\\
\subfigure[wormhole in de Sitter]
{\includegraphics[width=1.4 in]{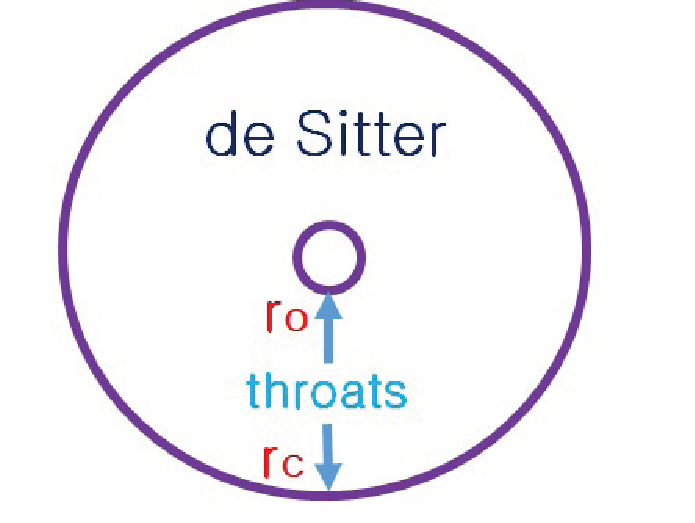}}
\subfigure[wormhole in anti-de Sitter]
{\includegraphics[width=1.4 in]{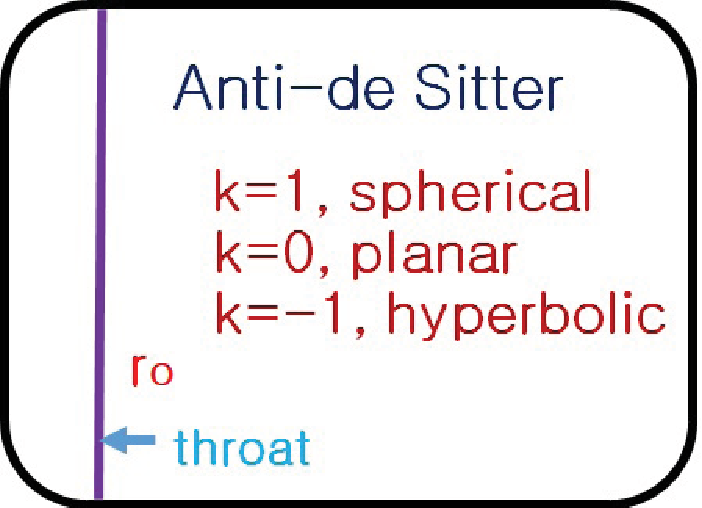}}
\end{center}
\caption{\footnotesize{(color online).
Conceptual diagram of the wormhole in de Sitter and anti-de Sitter spacetimes. }}
\label{conceptual}
\end{figure}
Figure \ref{conceptual} represents the conceptual diagram of the wormhole in de Sitter and anti-de Sitter spacetimes.

The wormhole throats investigated so far have finite area
and positive curvatures except for the planar thin-shell type~\cite{Lemos:2004vs}.
Consequently, it is natural to ask:
Is it possible to construct wormhole throats having negative or zero curvatures?
We demonstrate that it is indeed possible to construct a wormhole having a throat with alternative
topologies, and as a result, some throats may have infinite area.

\vspace{.3cm}
\textit{Charged static wormhole with a cosmological constant---} To construct charged wormholes with the cosmological constant,
we consider the action
\begin{equation}
I=\int d^4x \sqrt{-g}\Big[\frac{1}{16\pi}(R-2\Lambda-F_{\mu\nu}F^{\mu\nu})+{\cal L}_{\rm am}\Big] + I_{\rm b},
\end{equation}
where ${\cal L}_{\rm am}$ describes effective matter fields and $I_{\rm b}$ is the boundary term. We take the Newton constant $G=1$ for simplicity.
The cosmological constant $\Lambda \gtrless 0$ for de Sitter/anti-de Sitter spacetime.

The Einstein equation is
\begin{equation}
G_{\mu\nu}=R_{\mu\nu}-\frac{1}{2}R g_{\mu\nu} =8\pi T_{\mu\nu} -\Lambda g_{\mu\nu} \,, \label{einsteineq}
\end{equation}
where the stress-energy tensor takes the form
\begin{eqnarray}
T^{\mu\nu} - \frac{\Lambda}{8\pi} g^{\mu\nu} = T^{\mu\nu}_{\rm M} +  T^{\mu\nu}_{\rm am} - \frac{\Lambda}{8\pi} g^{\mu\nu} \,, \label{stente}
\end{eqnarray}
where $T^{\mu\nu}_{\rm M} = \frac{1}{4\pi}(F^{\mu}_{~\alpha}F^{\nu\alpha}- \frac{1}{4}g^{\mu\nu}F_{\alpha\beta}F^{\alpha\beta})$,
$T^{\mu\nu}_{\rm am} = (\varepsilon_{\rm am} + p_{t {\rm am}}) u^{\mu}u^{\nu} + p_{t{\rm am}} g^{\mu\nu}  + (p_{r {\rm am}} - p_{t {\rm am}}) x^{\mu}x^{\nu} $, $\varepsilon_{\rm am}$ is the energy density of the anisotropic matter, $u^{\mu}$ is four-velocity, and $x^{\mu}$ is a spacelike unit vector, respectively. The radial and the transverse (lateral) pressures are assumed to be linearly
proportional to the energy density:
\begin{equation}
p_{r {\rm am}}=w_{\rm 1}\varepsilon_{\rm am}\,,~~~ p_{t {\rm am}}=w_{\rm 2}\varepsilon_{\rm am}  \,, \label{eqofstate}
\end{equation}
then stress tensor for the anisotropic matter field in Eq.~(\ref{stente}) can be rewritten as~\cite{Kim:2024mam}
$T^{\mu}_{\nu {\rm am} }={\rm diag}(-\varepsilon_{\rm am}, w_1 \varepsilon_{\rm am}, w_2 \varepsilon_{\rm am}, w_2 \varepsilon_{\rm am})$.
The source-free Maxwell equations are given by
\begin{equation}
\nabla_{\nu}F^{\mu\nu} = \frac{1}{\sqrt{-g}}[\partial_{\nu}(\sqrt{-g}F^{\mu\nu})] =0 \,. \label{maxwelleq}
\end{equation}
We take the metric to be that of a static wormhole geometry
\begin{equation}
ds^2=  - f(r) dt^2 +  \frac{1}{g(r)} dr^2 + r^2 d\Sigma^2_k  \,. \label{stmetric}
\end{equation}
For the electrically   charged static geometry,
$F^{tr}=E^r=\sqrt{\frac{g(r)}{f(r)}} \frac{Q}{r^2}$ satisfy the source-free   Maxwell equations (\ref{maxwelleq}).
The electric field takes the form of $F^{{\hat a}{\hat b}}=e^{\hat a}_{\mu}e^{\hat b}_{\nu}F^{\mu\nu}$,
which gives $E^{\hat r}=\frac{Q}{r^2}$~\cite{Kim:2024mam}.

Then we take the metric functions as
\begin{equation}
f(r)=k+ \frac{Q^2}{r^2} -\frac{\Lambda}{3} r^2, \quad
g(r) = k + \frac{Q^2}{r^2} -\frac{\Lambda}{3} r^2  -\frac{b(r)}{r} \,,
\end{equation}
which denote the redshift function and include the wormhole shape function, respectively, as in Refs.~\cite{Kim:2001ri, Kim:2024mam},
in which $k=1$ for de Sitter spacetime, and $k=1$, $0$, or $-1$ for anti-de Sitter spacetime.
In the absence of matter, we assume that the metric function describes a topological (anti)-de Sitter spacetime~\cite{Mann:1996gj}.
Then, a two-dimensional surface with constants $t$ and $r$ has a positive, zero or
negative Gaussian curvature for $k=1$, $0$ or $-1$, respectively.
Accordingly, $d\Sigma^2_k$ denotes $d\Sigma^2_1 =d\theta^2 + \sin^2\theta d\phi^2$ for spherically symmetric one with $k=1$,
$d\Sigma^2_0 =dx^2 + dy^2$ for plane symmetric one with $k=0$,
and $d\Sigma^2_{-1} = d\psi^2 + \sinh^2\psi d\phi^2$ for hyperbolic symmetric one with $k=-1$, respectively.

The wormhole geometry attains its minimum radius at the typical throat $r=r_{o}$, satisfying $g(r_{o})=0$, i.e. $b(r_{o})=b_{o}$.
This condition yields $b_{o}=(k r^2_{o}+Q^2)/r_{o} - \frac{\Lambda}{3} r^3_o $,
where $b_o$ serves as a physical parameter of the wormhole for a given equation-of-state parameter $w_1$.
The region outside the wormhole throat satisfies $g(r)> 0$ and $f(r)> 0$.
The cosmological throat located at $r=r_{c} (> r_{o})$ shows the similar behaviors as the typical
throat with $b_{o}=(r^2_{c}+Q^2)/r_{c} - \frac{\Lambda}{3} r^3_c $ and $g(r_c)=0$.

The spatial area of the throat $r=r_{o}$ is given by $A_{r_o}=\int r^2_o d\Sigma^2_k$, hence the areal size of the throat
extends indefinitely for $k=0$ and $k=-1$.

We now consider Einstein equations.
The nonvanishing components of the Einstein tensor are given by
\begin{eqnarray}
G^{t}_{t}&&= - 8\pi \varepsilon = 8\pi( - \varepsilon_{c}  - \varepsilon_{\Lambda} - \varepsilon_{\rm am}) \nonumber \\
         &&= 8\pi(- \varepsilon_{c}) -\Lambda  -\frac{b'(r)}{r^2} \,,  \label{Etensor1} \\
G^{r}_{r}&&= 8\pi p_r = 8\pi(- \varepsilon_{c}  - \varepsilon_{\Lambda} + w_1 \varepsilon_{\rm am} ) \nonumber \\
         && = 8\pi( - \varepsilon_{c}) -\Lambda + \frac{3(Q^2-k r^2+\Lambda r^4)b(r)}{3r^3(Q^2+ k r^2)-\Lambda r^7} \,,   \label{Etensor2} \\
G^{\theta}_{\theta}&&= G^{\phi}_{\phi}= 8\pi p_t = 8\pi (\varepsilon_{c} - \varepsilon_{\Lambda}  + w_2 \varepsilon_{\rm am}) \nonumber \\
         && = 8\pi (\varepsilon_{c} ) -\Lambda  + \frac{3A(r)b(r) + B(r) b'(r) }{2r^3[3(Q^2+k r^2) - \Lambda r^4]^2}  \,,  \label{Etensor3}
\end{eqnarray}
where $A(r)=k r^2(\Lambda r^4 -9Q^2) + 3r^4 -6Q^4 + 10Q^2 \Lambda r^4 $, $B(r)= r^3(-3 k +2\Lambda r^2)[3(Q^2+k r^2) -\Lambda r^4]$,
$\varepsilon_c=\frac{Q^2}{8\pi r^4}$, $p_r$ is the radial pressure, $p_t$ is the transverse pressure,
and the prime denotes the derivative with respect to $r$.\footnote{For black holes with the metric function $f(r) = g(r)=k -\frac{2M}{r} + \frac{Q^2}{r^2} -\frac{\Lambda}{3} r^2 $, the Einstein equations are the same regardless of the value of $k$.}
When $\Lambda$ vanishes, only the $k=1$ case is permitted, and the corresponding geometry is reduced to
the case of charged wormholes with vanishing cosmological constant~~\cite{Kim:2024mam}.

From Eqs.\ (\ref{Etensor1}) and (\ref{Etensor2}), we obtain
\begin{equation}
b(r)=b_{o} \left(\frac{r b_{o}}{ k r^2+Q^2-\frac{\Lambda}{3}r^4 }\right)^{1/w_1}  \,,
\label{br}
\end{equation}
where $w_1$ is a constant.
Substituting this result back into Eqs.\ (\ref{Etensor1}) or (\ref{Etensor2}), we get
$\varepsilon_{\rm am}=- \frac{ k r^2-Q^2-\Lambda r^4}{8\pi w_1 r^4 }\left(\frac{b_{o} r}{ k r^2+Q^2-\frac{\Lambda}{3}r^4}\right)^{(w_1+1)/w_1}$.
After plugging those into Eq.\ (\ref{Etensor3}), we obtain
\begin{equation}
w_2(r)=-1 + \frac{(2Q^2 -k r^2)w_1}{k r^2 -Q^2 - \Lambda r^4} + \frac{3(2Q^2 + k r^2)(w_1 +1)}{6(Q^2 + k r^2) -2 \Lambda r^4} \,.
\label{eqodst2}
\end{equation}
Then, the energy density and pressures are given by
\begin{eqnarray}
\varepsilon&=&\frac{Q^2}{8\pi r^4} + \frac{\Lambda}{8\pi} +\varepsilon_{\rm am} \,, \nonumber \\
p_r &=& -\frac{Q^2}{8\pi r^4} -\frac{\Lambda}{8\pi}  + w_1 \varepsilon_{\rm am}\,, \\
p_t &=&  \frac{Q^2}{8\pi r^4}  -\frac{\Lambda}{8\pi}  + w_2 \varepsilon_{\rm am}\,. \nonumber
\label{edenpre}
\end{eqnarray}

Now let us describe the conditions that the above solution describes a wormhole geometry.

We first check the flare-out condition and the energy condition~\cite{Morris:1988cz, Hochberg:1997wp}.
These conditions must be satisfied by the matter supporting the wormhole near the wormhole throat,
and this investigation places constraints on the parameters of the equation of state for the matter supporting the wormhole,
in addition to the amount of charge and the cosmological constant.
The flare-out condition through the embedding geometry at $t=\rm const.$ and $\theta= \rm const.$
($\sin\theta=1$ and $\sinh\psi=1$) becomes
\begin{equation}
\label{flare-out}
\frac{d^2 r}{dz^2} = \frac{r[r(b(r)-rb'(r))-2Q^2 -\frac{2\Lambda}{3} r^4]}{2[k rb(r) - k(Q^2 - \frac{\Lambda}{3} r^4 + k r^2) +r^2]^2} > 0 \,,
\end{equation}
where $z$ denotes the coordinate in the embedding space. Thus $b(r)$ around the throat governs the  flare-out condition.

Substituting Eq.~(\ref{br}) into the flare-out condition, we get the numerator at the throat
\begin{equation}
\label{exoticity22}
N(r_o) = (k r^2_{o}-Q^2-\Lambda r^4_o)\left(1+1/w_1 \right)  > 0 \,.
\end{equation}
We choose $w_1 > 0$ or $w_1 < -1$ as the expanding case with the cosmological constant~\cite{Kim:2024mam}.
Conversely,
if $-1 < w < 0$, the contribution of the $b(r)$ increases, resulting in improper behaviors for the metric functions,
$g(r) < 0$ in most of the region where $f(r) > 0$.

Now we want to understand the relation among the size of the wormhole throat and $Q$, $\Lambda$, and $k$ to satisfy the flare-out condition.
The typical throat has the minimum radius, thus it satisfies $ \frac{1- \sqrt{1- 4\Lambda Q^2}}{2\Lambda} < r^2_o <  \frac{1+ \sqrt{1- 4\Lambda Q^2}}{2\Lambda}$
for the de Sitter spacetime, in which $\Lambda Q^2 < 1/4$.
The cosmological throat has a maximum radius $r_c$, so the inequality in Eq.~(\ref{flare-out}) is reversed (i.e., becomes negative) because $r$ decreases as
$z$ increases.
That is to say $N(r_c)<0$, it gives $r^2_c < \frac{1- \sqrt{1- 4\Lambda Q^2}}{2\Lambda} $
or  $r^2_c > \frac{1+ \sqrt{1- 4\Lambda Q^2}}{2\Lambda} $, and we take the latter.
Equation~(\ref{exoticity22}) gives $r^2_o > \frac{-k + \sqrt{k^2+ 4|\Lambda| Q^2}}{2|\Lambda|}$ for the anti-de Sitter spacetime.
For vanishing $\Lambda$ with $k=1$, it reduces $r^2_o > Q^2$ and $w_1 > 0$ or $w_1 < -1$~\cite{Kim:2024mam}.

One can introduce the exoticity function~\cite{Morris:1988cz} as
\begin{eqnarray}
&&\zeta(r) \equiv \frac{-p_r - \varepsilon}{|\varepsilon|}  \nonumber \\
&& = \frac{(k r^2-Q^2 -\Lambda r^4 )(1+w_1)\left( \frac{r b_o}{k r^2+Q^2 -\frac{\Lambda}{3}r^4} \right)^{(w_1+1)/w_1}}{8\pi r^4 w_1 |\varepsilon|}\,.
\label{exoticity}
\end{eqnarray}
When the exoticity function is positive, the null energy condition is violated.
At the both throats, this one turns out to be
\begin{eqnarray}
\zeta(r_0)=\frac{(k r^2_o-Q^2-\Lambda r^4_o)(1+1/w_1)}{8\pi r^4_o |\varepsilon(r_o)|} \,,
\label{exoticity3}
\end{eqnarray}
where this one takes the same form as (\ref{exoticity22}) up to a positive definite multiplication factor.
Thus, the wormhole supported by that matter could satisfy the flare-out condition.
We note that the null energy condition is violated at the typical throat of a wormhole,
whereas it is not violated at the cosmological throat.

Now we check the Petrov type of our wormhole solution.
The Petrov classification corresponds to the scheme of classification based on the Weyl conformal tensor.
It describes the possible algebraic symmetries of the Weyl tensor~\cite{Petrov:2000bs, Newman:1961qr, Carmeli:2001ay}.
The appropriate combinations for classification are known as the five complex tetrad components of the Weyl conformal tensor:
\begin{eqnarray}
\label{fcwct}
\Psi_0 &\equiv& C_{\mu\nu\alpha\beta} l^{\mu} m^{\nu} l^{\alpha} m^{\beta} \,, \nonumber \\
\Psi_1 &\equiv& C_{\mu\nu\alpha\beta} l^{\mu} n^{\nu} l^{\alpha} m^{\beta} \,, \nonumber \\
\Psi_2 &\equiv& C_{\mu\nu\alpha\beta} l^{\mu} m^{\nu} {\bar m}^{\alpha} n^{\beta} \,,  \\
\Psi_3 &\equiv& C_{\mu\nu\alpha\beta} l^{\mu} n^{\nu} {\bar m}^{\alpha} n^{\beta} \,, \nonumber \\
\Psi_4 &\equiv& C_{\mu\nu\alpha\beta} n^{\mu} {\bar m}^{\nu} n^{\alpha} {\bar m}^{\beta} \,, \nonumber
\end{eqnarray}
where the corresponding null vectors are given from (\ref{stmetric})
\begin{eqnarray}
\label{nullvetorll}
l^{\mu} &=&  \frac{1}{\sqrt 2} \left( \frac{1}{\sqrt f(r)} \delta^{\mu}_0 + {\sqrt g(r)} \delta^{\mu}_1  \right) \,, \nonumber \\
n^{\mu} &=& \frac{1}{\sqrt 2} \left( \frac{1}{\sqrt f(r)} \delta^{\mu}_0 - {\sqrt g(r)}  \delta^{\mu}_1  \right) \,, \nonumber \\
m^{\mu} &=& \frac{1}{\sqrt{2} r} \left( \delta^{\mu}_2 + i_k  \delta^{\mu}_3  \right) \,,  ~
{\bar m}^{\mu} = \frac{1}{\sqrt{2} r} \left( \delta^{\mu}_2 - i_k  \delta^{\mu}_3 \right) \,,
\end{eqnarray}
where $i_{1} = \frac{i}{\sin\theta}$, $i_0 = i$, and $i_{-1}= \frac{i}{\sinh\theta}$.
Explicit calculation in the Appendix of Eq~(\ref{fcwct}) shows that only
$\Psi_2$ is nonzero
\begin{eqnarray}
\Psi_2 &\equiv& C_{\mu\nu\alpha\beta} l^{\mu} m^{\nu} {\bar m}^{\alpha} n^{\beta}  \nonumber \\
&=& \frac{J(r) + K(r) }{24 r^2 f^2(r)} \,,
\label{fcwct04}
\end{eqnarray}
where $J(r) = -r^2 g(r) f'^2(r) + f^2(r) [4g(r) -2r g'(r)-4]$ and $K(r) = r f(r) [f'(r) (r g'(r)-2g(r)) + 2r g(r) f''(r) ] $
When $g(r)=f(r)$ it reduces to $\frac{r^2 f''(r) - 2r f'(r) + 2f(r) -2}{12 r^2}$.
For example, $\Psi_2=\frac{Q^2-Mr +\frac{r^2}{6} (k-1)}{r^4}$ for the charged black hole, in which we used $f(r)=k - \frac{2M}{r} + \frac{Q^2}{r^2}- \frac{\Lambda}{3} r^2$.

Thus, this geometry corresponds to the Type $D$, in which there are two double principal null directions
that correspond to radially ingoing and outgoing null rays near the wormhole, in which
$l^{\mu}$ and $n^{\nu}$ denote the outgoing and ingoing null vector, respectively.

\vspace{.3cm}
\textit{Radial geodesics ---} We now consider radial geodesics for both the light and massive particles.
One can take  $m\neq0$ for timelike geodesics and $m = 0$ for null geodesics~\cite{Carter:1968rr, Perlick:2003vg, Jeong:2023hom}.
The radial geodesic equations are given by
\begin{eqnarray}
\label{raged}
\left(\frac{dr}{d\lambda} \right)^2  +   V_{\rm eff} =0 \,, ~~
\end{eqnarray}
where
\begin{equation}
V_{\rm eff} =
\left\{
\begin{array}{cc}
 g(r)\left[ - \frac{E^2}{f(r)} + \frac{L^2_z}{r^2} \right] &  m=0  \\
\frac{g(r)}{m^2}\left[ m^2 - \frac{\left(E\mp \frac{eQ}{r} \right)^2}{f(r)} + \frac{L^2_z}{r^2} \right]  & m\neq 0
\end{array}
\right. .
\end{equation}
We take $\sin\theta =1$ and $\sinh\theta =1$.
If one chooses $+$ signature in one universe, one needs to choose $-$ signature in the other universe
because of the electric field heading opposite direction.
\begin{figure}[h]
\begin{center}
\subfigure[light in de Sitter]
{\includegraphics[width=1.4 in]{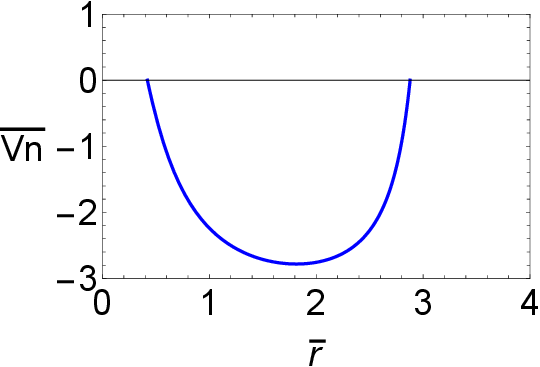}}
\subfigure[massive particle in de Sitter]
{\includegraphics[width=1.4 in]{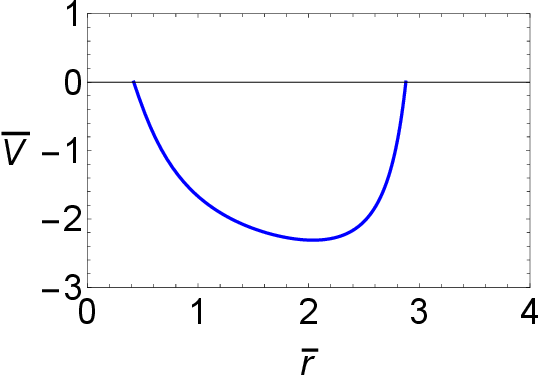}}
\subfigure[light in anti-de Sitter]
{\includegraphics[width=1.4 in]{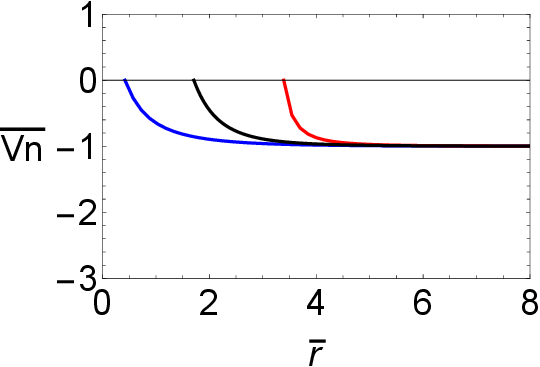}}
\subfigure[massive particle in anti-de Sitter]
{\includegraphics[width=1.4 in]{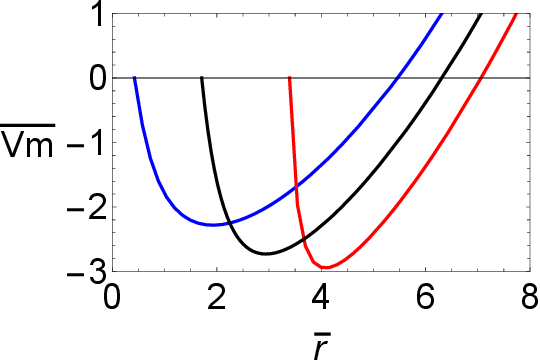}}
\end{center}
\caption{\footnotesize{(color online).
The shape of the effective potential
for the radial geodesics in de Sitter and anti-de Sitter spacetimes.}}
\label{veffdeantide}
\end{figure}
Figures \ref{veffdeantide}~(a) and (b) represent the  characteristic forms of $V_{\rm eff}$ for radial null and timelike geodesics
in de Sitter spacetime, respectively.
There exist two kinds of wormhole throat, a typical/cosmological throat determined by $g(r)=0$.
Thus particles or lights that come into our universe through the wormhole
throat proceed and plunge into the cosmological throat.

Figures \ref{veffdeantide}~(c) and (d) represent the characteristic forms of $V_{\rm eff}$
for radial null and timelike geodesics in anti-de Sitter spacetime, respectively.
The blue line is for $k=1$, the  black line for $k=0$, while the red line for $k=-1$.
There are one/two roots satisfying $V_{\rm eff}=0$ for null/timelike geodesics, respectively. The left roots are from
$g(r)=0$ at the wormhole throat, as shown in Eq.~(\ref{raged}).
The effective potential for the null ray is always less than zero outside of the throat,
i.e., light comes out of the wormhole throat and proceeds into infinity as shown in Fig.~\ref{veffdeantide}$(c)$.
While, for the timelike geodesics, the effective potential has two roots, as illustrated in Fig.~\ref{veffdeantide}$(d)$.
One of the points corresponds to $g(r)=0$ and the other to the square bracket quantity equals zero.
Due to the attractive nature of anti-de Sitter spacetime, a potential barrier forms,
causing a massive particle to bounce off at the right root.
In this way, the particle will oscillate through the wormhole throat between both universes.
\vspace{.3cm}

\textit{Conclusions ---}
We have constructed charged, traversable wormhole geometries by solving the Einstein-Maxwell equations in the presence of a cosmological constant. In de Sitter spacetime, two types of wormhole throats may exist--referred to as the typical and cosmological throats--located at different radial positions.
In contrast, only the typical throat is permitted in anti-de Sitter spacetime.
Notably, the throat can support three distinct topologies, and in two of these cases, the area of the throat can go to infinite.
What are the implications of a wormhole throat with infinite area?
An observer moving radially toward such a throat must traverse it.
However, orbital motion around the wormhole is no longer possible in these configurations because of the infinite area.

In de Sitter wormhole spacetimes, the behaviors of radial null and timelike geodesics are qualitatively similar: a particle or light ray passes through the wormhole, entering from one throat (either typical or cosmological) and emerging from the other.
In contrast, in anti-de Sitter spacetime, null geodesics can pass through the throat and escape to spatial infinity.
However, massive particles are subject to a potential barrier at large $r$, induced by the negative cosmological constant, which prevents them from escaping.
As a result, they undergo oscillatory motion between the two asymptotic regions, cyclically traversing the wormhole.

In this work, we present new families of traversable wormhole solutions that exist in spacetimes with a cosmological constant, consistent with the Einstein-Maxwell equations. These solutions are expected to reveal physical phenomena distinct from those arising in conventional wormhole geometries.

\vspace{.3cm}
\textit{Acknowledgments ---} We are grateful to Miok Park and Yun Soo Myung for their hospitality during SGC $2024$,
which was held at IBS, as well as to Jeong-Hyuck Park, Minkyoo Kim, and Wontae Kim for their valuable discussions
at the $3$rd Han-Gang Gravity Workshop.
We express our gratitude to Sung-Won Kim and Bum-Hoon Lee for consistently encouraging us to delve into wormhole physics.
H.-C. Kim (RS-2023-00208047) and W. Lee (RS-2022-NR075087, CQUeST: RS-2020-NR049598) were supported by Basic Science
Research Program through the National Research Foundation of Korea funded by the Ministry of Education.


\vspace{.3cm}
\textit{Appendix ---} In $4$-dimensional spacetime, the Weyl conformal tensor is given by~\cite{Carmeli:2001ay}
\begin{eqnarray}
&&C_{\mu\nu\alpha\beta} = R_{\mu\nu\alpha\beta} + \frac{R}{6} (g_{\mu\alpha} g_{\nu\beta} - g_{\mu\beta} g_{\nu\alpha}) R  \nonumber \\
&& - \frac{1}{2} (g_{\mu\alpha}R_{\nu\beta} - g_{\mu\beta}R_{\nu\alpha} - g_{\nu\alpha}R_{\mu\beta}
                          + g_{\nu\beta}R_{\mu\alpha}  ) \,.
\label{wcten}
\end{eqnarray}
This tensor has the same symmetry properties as the Riemann curvature tensor and non-vanishing components:
\begin{eqnarray}
\label{nvacow}
&&C_{0101}  = \frac{f(r)}{3r^2} \left(1- \frac{1}{g(r)} \right)
          -\frac{f'(r)}{6} \left(\frac{1}{r} + \frac{f'(r)}{2f(r)} \right) \nonumber \\
          && - \frac{g'(r)}{6g(r)} \left(\frac{f(r)}{r} - \frac{f'(r)}{2} \right) +\frac{f''(r)}{6}  \,, \nonumber \\
&& C_{0202} = \frac{f(r)}{6} \left(1- g(r) \right)
          +\frac{r g(r) f'(r)}{12}  \left(1 + \frac{r f'(r)}{2f(r)} \right) \nonumber \\
          && +\frac{r g'(r)}{12} \left(f(r) - \frac{r f'(r)}{2} \right) - \frac{r^2 g(r) f''(r) }{12} \,, \nonumber \\
&&C_{0303} =  C_{0202} \left( S^2_k  \right)    \,,  \\
&&C_{1212} =  C_{0101} \left( \frac{r^2}{2f(r)} \right) \,, \nonumber \\
&&C_{1313} =  C_{1212} \left( S^2_k \right)  =  C_{0101} \left( \frac{r^2 S^2_k}{2f(r)} \right)    \,,  \nonumber \\
&&C_{2323} =  C_{0303} \left( \frac{2r^2}{f(r)}  \right) =  C_{0202} \left( \frac{2r^2 S^2_k}{f(r)}  \right)    \,,  \nonumber
\end{eqnarray}
where $S^2_{1} = \sin^2\theta $, $S^2_{0} = 1 $, and $S^2_{-1} = \sinh^2\theta $.

From Eqs.~(\ref{nullvetorll}) and (\ref{nvacow}), non-vanishing components are given by
\begin{eqnarray}
\label{nvacopsi01}
&& C_{1313} l^{1} m^{3} l^{1} m^{3}  = - C_{1212}l^{1} m^{2} l^{1} m^{2}  \,, \nonumber \\
&& C_{0303} l^{0} m^{3} l^{0} m^{3} = - C_{0202}l^{0} m^{2} l^{0} m^{2} \,, \nonumber \\
&& C_{0303} n^{0} {\bar m}^{3} n^{0} {\bar m}^{3}  = - C_{0202} n^{0} {\bar m}^{2} n^{0} {\bar m}^{2} \,,  \\
&& C_{1313} n^{1} {\bar m}^{3} n^{1} {\bar m}^{3}   = - C_{1212} n^{1} {\bar m}^{2} n^{1} {\bar m}^{2} \,, \nonumber \\
&& C_{0330} l^{0} m^{3} {\bar m}^{3} n^{0}  = C_{0220} l^{0} m^{2} {\bar m}^{2} n^{0}  \,, \nonumber \\
&& C_{1331} l^{1} m^{3} {\bar m}^{3} n^{1} = C_{1221} l^{1} m^{2} {\bar m}^{2} n^{1} \,. \nonumber
\end{eqnarray}


\end{document}